\documentclass[12pt, letterpaper]{article}
\usepackage[utf8]{inputenc}
\usepackage{subfig}
\usepackage{graphicx}
\usepackage{amsmath, amsthm, amsfonts}
\usepackage[numbers]{natbib}
\usepackage{algorithm, algorithmic}
\usepackage{placeins}
\usepackage{rotating}
\usepackage{booktabs}
\usepackage{multirow} 
\usepackage[font=small,labelfont=bf,tableposition=top]{caption}
\usepackage{xfrac}
\usepackage{mathrsfs}
\usepackage{lineno}
\usepackage{layouts}
\usepackage{url}

\usepackage[text={16cm,23cm},centering]{geometry}

\begin{document}

\title{ChainSQL: A Blockchain Database Application Platform  
}

\footnotesize\date{\today}

\author{
 Muhammad Muzammal, Qiang Qu, Bulat Nasrulin, Anders Skovsgaard$^{\ddag}$\\
     \footnotesize Shenzhen Institutes of Advanced Technology\\
     \footnotesize Chinese Academy of Sciences, China\\
     \footnotesize $^{\ddag}$2600 Security, Denmark\\
     \footnotesize 
       \{muzammal, qiang, bulat\}@siat.ac.cn, anders@2600.dk\\
}

\maketitle
\begin{abstract}
A blockchain is a decentralised linked data structure that is characterised by its inherent resistance to data modification, but it is deficient in search queries, primarily due to its inferior data formatting. A distributed database is also a decentralised data structure which features quick query processing and well-designed data formatting but suffers from data reliability. In this demonstration, we showcase a blockchain database application platform developed by integrating the blockchain with the database, i.e. we demonstrate a system that has the decentralised, distributed and audibility features of the blockchain and quick query processing and well-designed data structure of the distributed databases. The system features a tamper-resistant, consistent and cost-effective multi-active database and an effective and reliable data-level disaster recovery backup. The system is demonstrated in practice as a multi-active database along with the data-level disaster recovery backup feature.
\end{abstract}


\section{Introduction}

Digital or crypto currencies, for example, Bitcoin\footnote{https://bitcoin.org/en} and  Ethereum\footnote{https://www.ethereum.org}, have recently witnessed a tremendous interest from the user~\cite{cointelegraph} as well as the developer community~\cite{rosa, DBLP:journals/corr/abs-1708-05665}. The crypto currencies are essentially smart contracts between the users which are executed using a data structure referred to as `blockchain'. Thus, a blockchain stores financial transactions whilst satisfying the following two constraints: (a) anyone should be able to write to the blockchain, and (b) there should not be any centralised control. 

A blockchain is a database and an application software on top of it~\cite{8029983} that dictates the data definition and data update mechanism for the blockchain. A blockchain does not only allows to add new data to the database but it also ensures that all the users on the network have exactly the same data. Thus, a blockchain is a distributed and decentralised linked data structure for data storage and retrieval which also ensures that the data is decentralised and is resistant to data modifications.

One of the limitations of blockchain is its inherent deficiency in search query processing~\cite{wright_filippi_2015} primarily due to the linked data storage and the absence of a well-defined data indexing structure for various queries. Bitcoin, for instance, is the most notable blockchain network, however in practice it has two limitations: (a) it takes a considerable amount of time, possibly up to ten (10) minutes, for a transaction to be issued and verified and the final confirmation may take up to an hour, and (b) a new block can only be generated by miners, which requires extensive computations.

Databases, in addition to having a defined data structure are optimised for faster query processing~\cite{DBLP:conf/sigmod/HadianNMQ16,DBLP:conf/fskd/QuQSW08,routing,DBLP:journals/debu/QuCJS15,DBLP:conf/ssdbm/QuLYJ14,DBLP:journals/tkde/QuLZJ16}; but are not resistant to data modifications~\cite{PHILIPCHEN2014314,chainsqlfgcs,DBLP:conf/mdm/NasrulinMQ18}. More specifically, distributed databases have the following limitations: (a) database can be tampered either by a malicious user or by the database administrator~\cite{DBLP:journals/access/MuzammalGRQAJ18,DBLP:journals/tbd/LiuQ0N15,DBLP:journals/ijgi/HasanQLCJ18}, (b) the backup-based disaster recovery scheme of the database cannot be normally activated in the event of a system failure caused by data loss, and (c) the multiple copies of the database are not entirely consistent and the data synchronisation operations are required to resolve data conflicts. 

Therefore, a database system is desirable that has the features of the blockchain and the distributed databases combined together such that the inherent resistance of the blockchain to data modification and the query speed of the distributed databases is achieved. In this work, we showcase \textsc{ChainSQL}\footnote{The source code for \textsc{ChainSQL} is available online at: https://github.com/ChainSQL/chainsqld}, a new blockchain-based log database system that combines the features of the blockchain technology and the distributed databases.  
%
\begin{figure}
	\centering 
		\subfloat[Direct blockchain write access]{\includegraphics[scale=0.55]{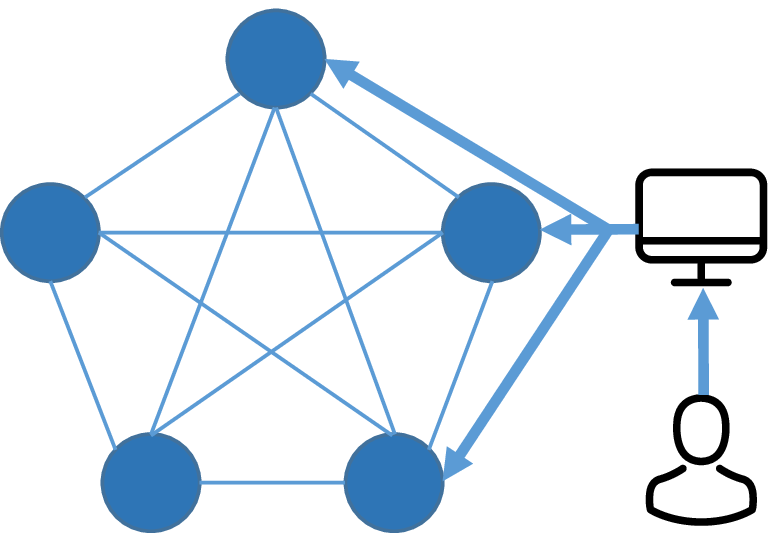}}\quad
		\subfloat[Fast database read access]{\includegraphics[scale=0.55]{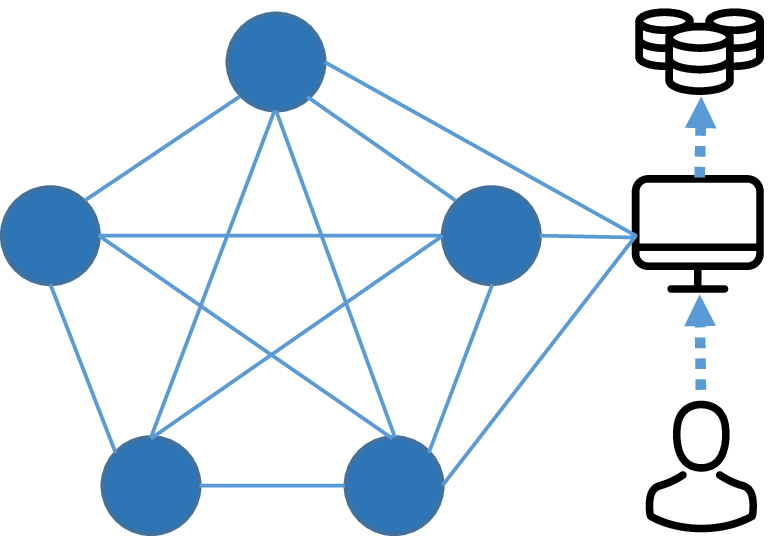}}\quad
		\subfloat[Fast read-database/write-blockchain access]{\includegraphics[scale=0.55]{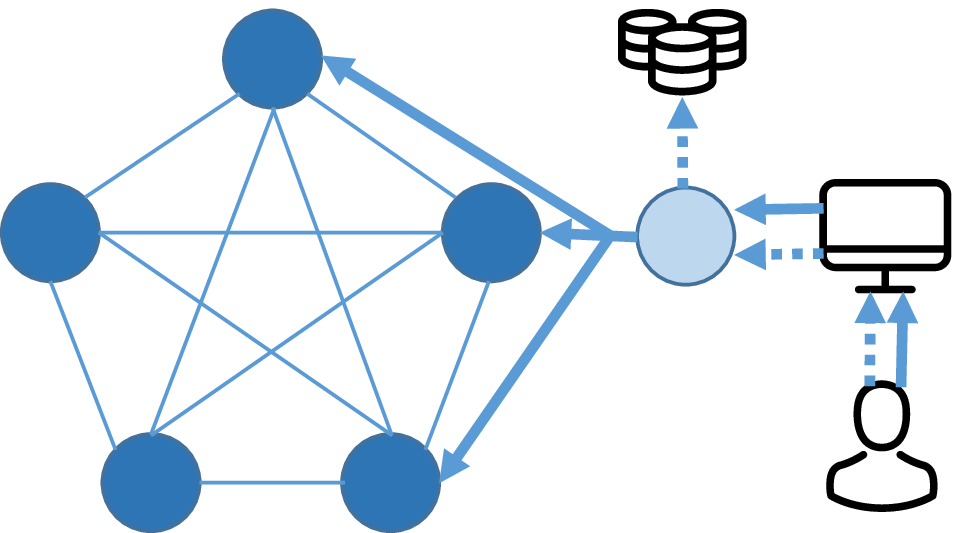}}
	\caption{An overview of the \textsc{ChainSQL} access mechanism; the blockchain network is accessed (a) directly by the client for write operations, (b) by using an underlying database for read operations, or (c) by configuring a database on a blockchain node for read-database/write-blockchain operations.}
    \label{fig:1}
\end{figure}

\textsc{ChainSQL} has two usecases, each of which is implemented as a middleware between the enterprise application and the underlying database: (i) a multi-active database middleware that connects the enterprise application with the database system, and (ii) a database disaster recovery middleware that connects the database production nodes with the disaster recovery nodes. 
Details about the usecases are provided in Section~\ref{demo}.

\begin{figure}[h]
	\begin{center}
		\begin{tabular}{c}
			\includegraphics[scale=.85]{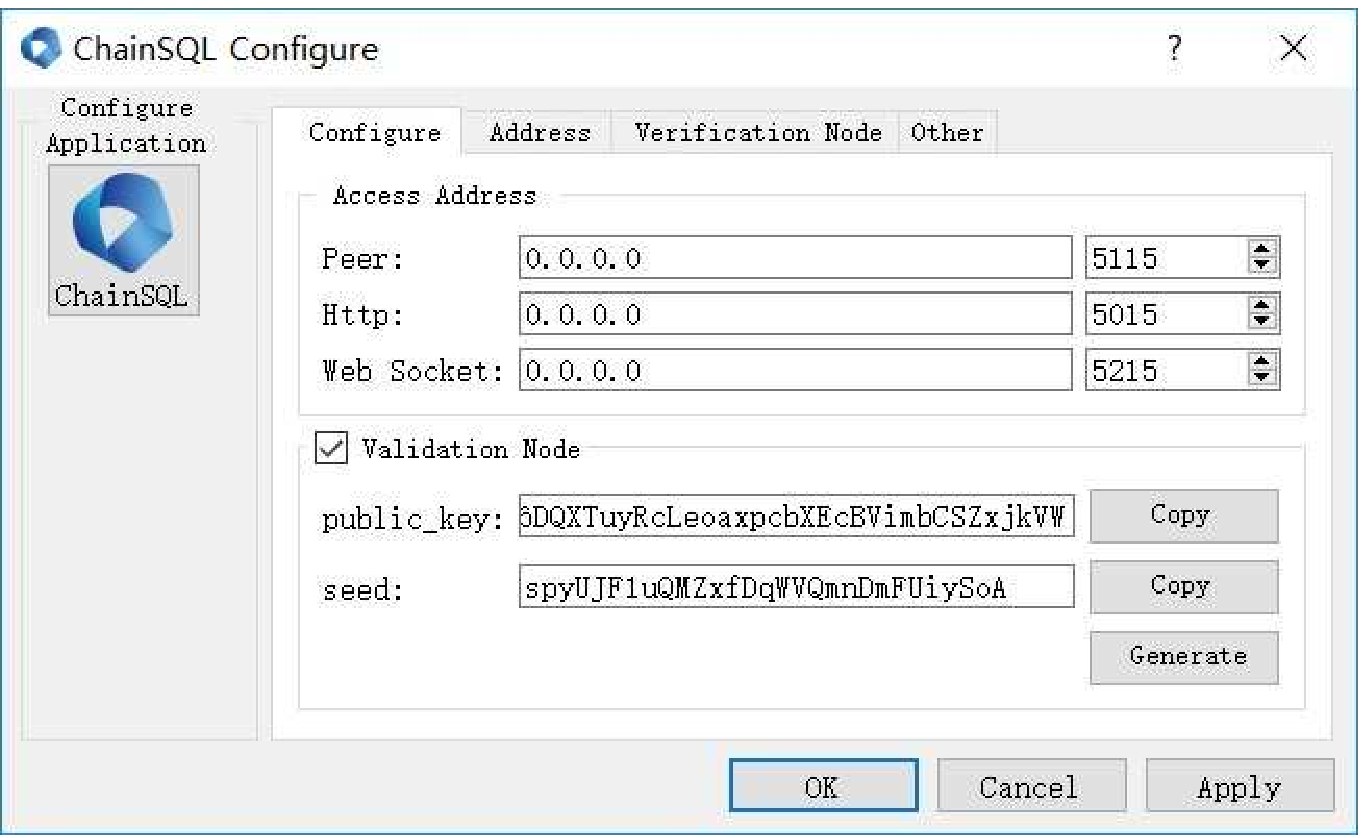} \\
			
		\end{tabular}
	\end{center}
    \caption{\textsc{ChainSQL} configuration interface.}
    \label{fig:6}
\end{figure}

\textit{\textsc{ChainSQL} Highlights.} \textsc{ChainSQL} features a secure design due to the authorisation requirement to access the personal user data. The transactions are stored in the blockchain whereas the actual data is stored in the database. The data is distributed to improve service availability. Many-to-one disaster recovery architecture allows a single backup centre to be used with multiple production sites. The backup database can be operated without data recovery. Thus, \textsc{ChainSQL} not only provides the instantaneity of the traditional databases but also the security of the blockchain~\cite{DBLP:journals/tbd/LiuQ0N15}. It can be easily configured with the commonly used databases (see Figure~\ref{fig:6} for the configuration interface) such as MySQL, Oracle, IBM DB2, and it is easy to program using APIs. As the database log is immutable, the history database actions are preserved and therefore, it allows auditing using the data stored in the blockchain. The integration of the blockchain with the new applications is simple as it only requires using the \textsc{ChainSQL} interface rather than the database interface.
\section{ChainSQL Overview}
In this section, we present an overview of \textsc{ChainSQL} as the context for understanding the applications that we demonstrate in Section~\ref{demo}. \textsc{ChainSQL} regards three aspects, a blockchain network, a database and a set of users, an outline of which is given as follows: 
\begin{enumerate}
    \item The blockchain network used by \textsc{ChainSQL} is Ripple\footnote{https://ripple.com} due to the following reasons: (a) Ripple is able to issue and verify transactions quickly (within four (04) seconds); (b) It avoids extensive Bitcoin like computations for new block generation by incorporating its own 'unique node list' (UNL) scheme. More details about blockchain consensus models and Ripple can be found at the study~\cite{baliga2017}.
    \item A database is configured on top of the blockchain nodes which is synchronised with the blockchain and facilitates quick database like blockchain `read' operations.
    \item Users can query the \textsc{ChainSQL} network as follows: (a) directly query the blockchain network (b) create a database blockchain node for fast access, or (c) both (a) and (b) combined together. An overview of \textsc{ChainSQL} access mechanism is shown in Figure~\ref{fig:1}.
\end{enumerate}
We now briefly discuss \textsc{ChainSQL} components:
 
\textit{Application Interface.} The access to the blockchain is via APIs provided as an interface to the application and therefore, a transaction command to the blockchain is similar to a database operation in the user context. Multiple programming languages are supported by \textsc{ChainSQL} APIs\footnote{https://github.com/ChainSQL}, which enable the flexibility and applicability of \textsc{ChainSQL}. 

\textit{Database Operations.} The database operations are performed in a real-time environment. The blockchain network directly transmits the transaction data to the corresponding database for processing. The consensus mechanism to authenticate the validity of a transaction is given as follows: a transaction is authenticated by a set of nodes that are a subset of the blockchain network nodes drawn by the implementation of a UNL scheme. If a transaction is authenticated, it is sent to the blockchain network for consensus and is subsequently written into the database. If consensus fails, the database operation is rolled back. The entire process is completed within a few seconds and therefore, the user is updated in near real-time about the transaction status.   

\textit{Database Recovery.} One of the blockchain network nodes is configured with a database to keep transactions in the blockchain or to execute database operations to recreate a new table. A node on the blockchain network can be either a full-record node (that stores all the transactions on the blockchain network) or a partial-record node.
\section{Demonstration}
\label{demo}
We demonstrate two usecases of \textsc{ChainSQL}: (i) a multi-active database middleware for connecting the user application with the underlying database, and (ii) a disaster recovery middleware that connects user application production nodes with the disaster recovery nodes. 

%
%
%
\begin{figure}[ht]
	\begin{center}
		\begin{tabular}{c}
			\includegraphics[scale=.35]{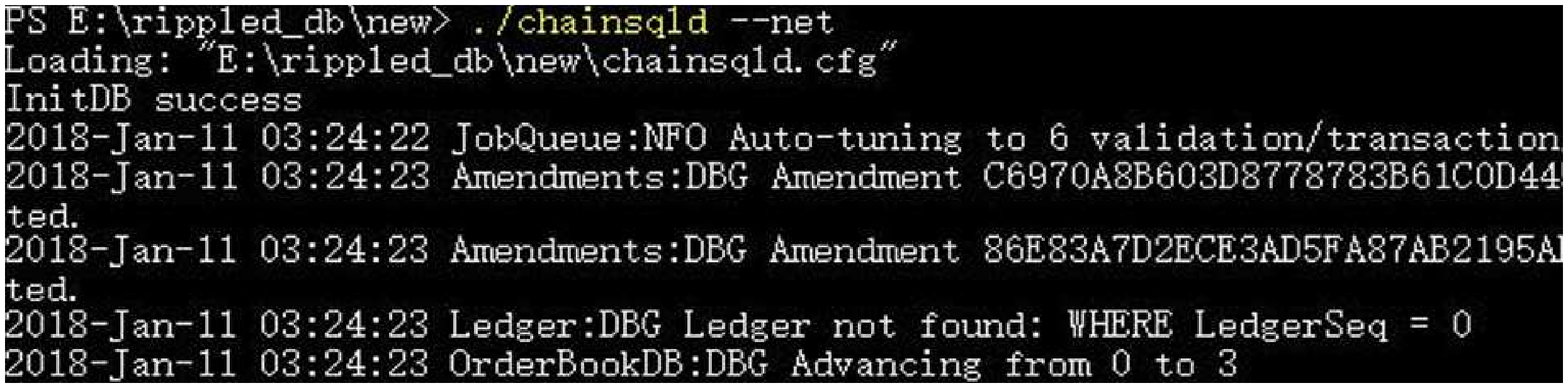} \\
			(i) \\
			\includegraphics[scale=.36]{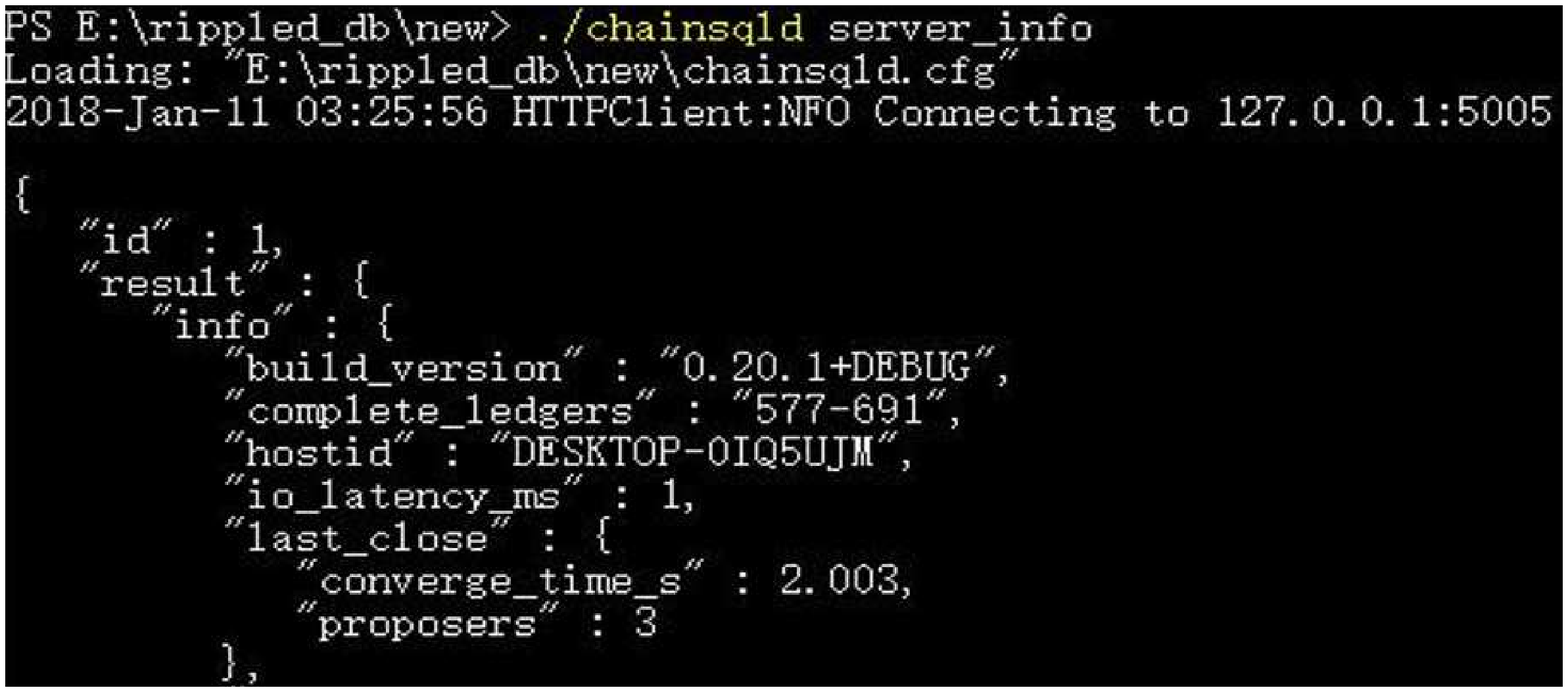}\\
			(ii) \\
			\includegraphics[scale=.24]{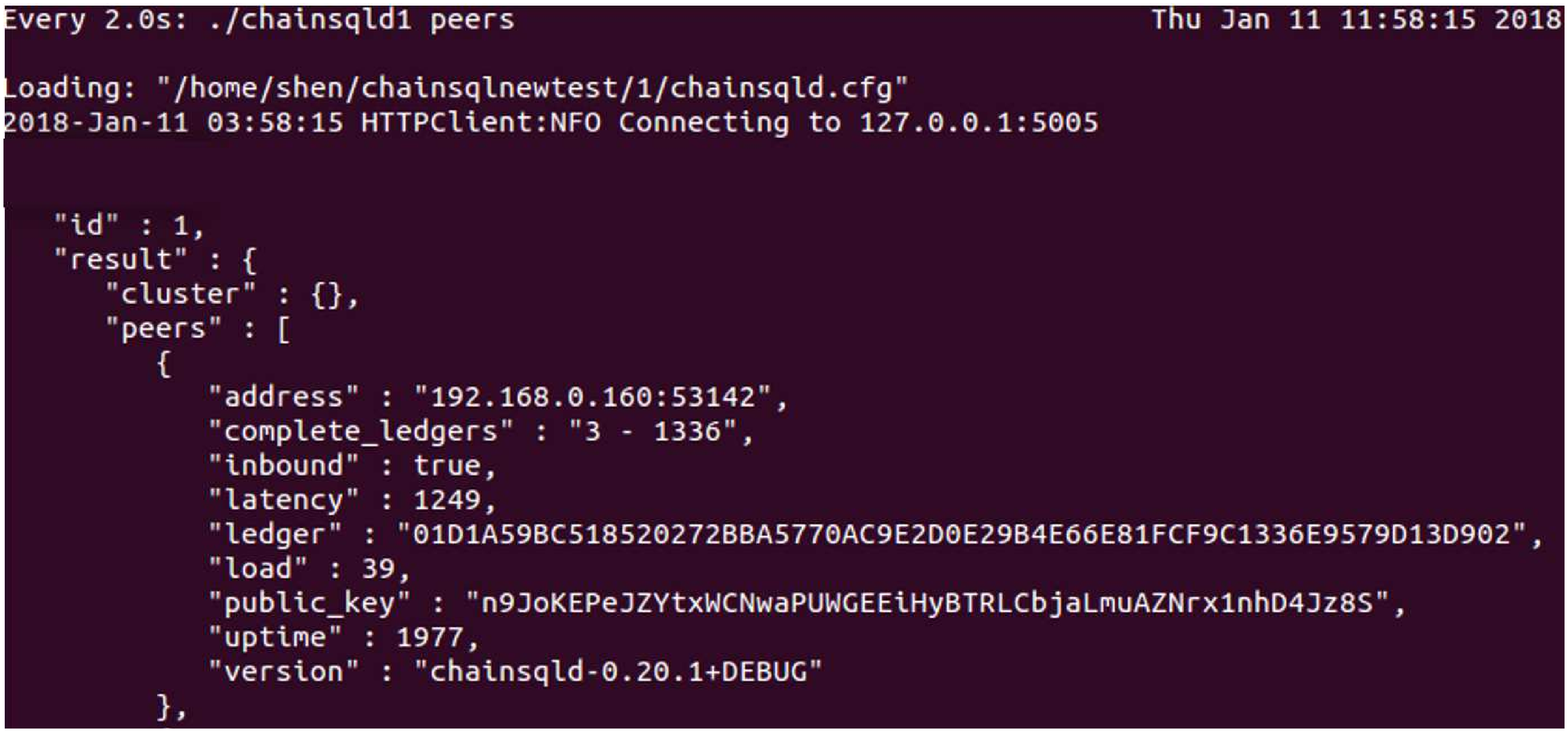}\\
			(iii) \\
			
		\end{tabular}
	\end{center}
    \caption{\textsc{ChainSQL} initialisation process. (i) Start the database, (ii) view the server information, and (iii) view \textsc{ChainSQL} peers.}
    \label{fig:7}
\end{figure}
%
%
%
%
%
%
\begin{figure}[ht]
	\begin{center}
		\begin{tabular}{c}
			\includegraphics[scale=.65]{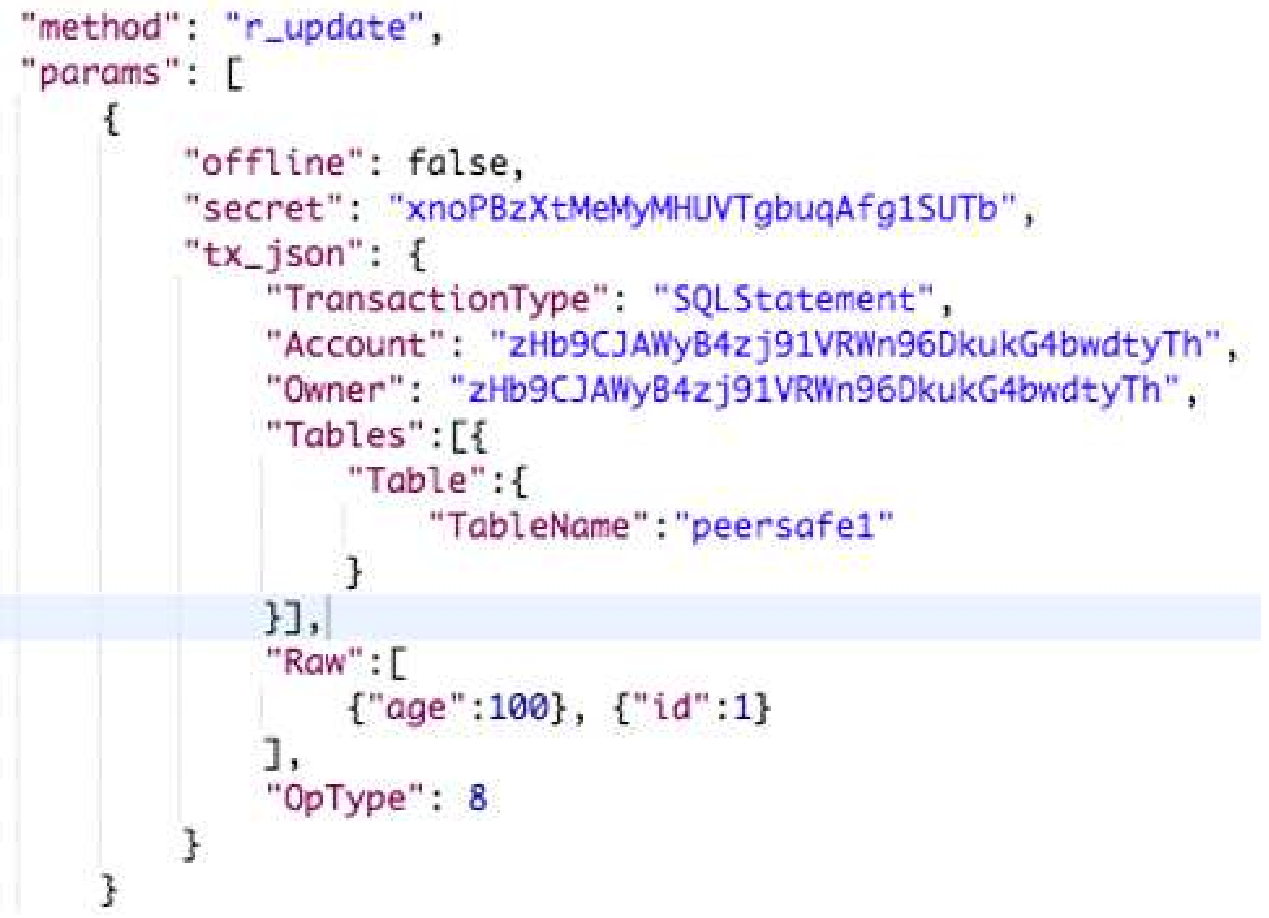} \\
			(i) \\
			\includegraphics[scale=.5]{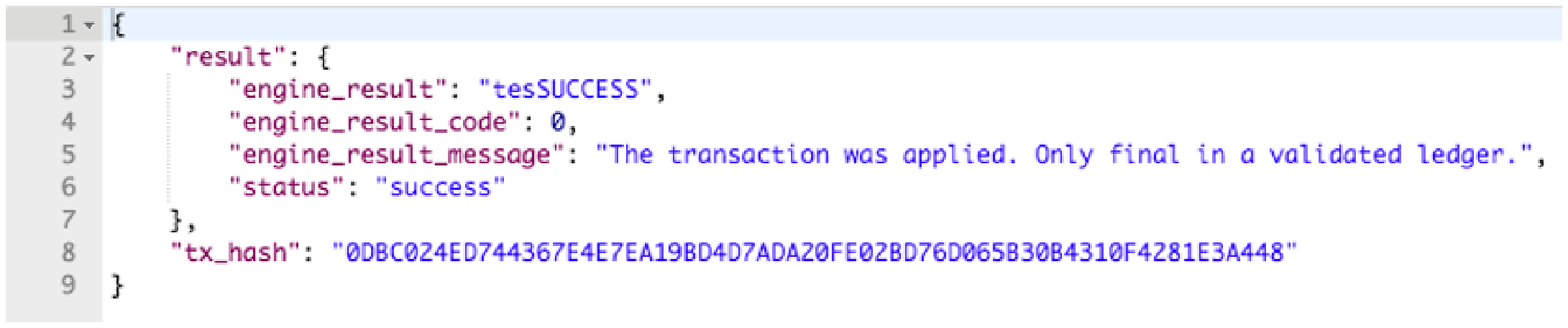}\\
			(ii) \\
		\end{tabular}
	\end{center}
\vspace{-10px}
    \caption{A sample of the \textsc{ChainSQL} Update operation. (i) Update query, and (ii) query execution result.}
    \label{fig:8}
\end{figure}
%
%
%
%
	
%
%
%
%
%
\subsection{Multi-active Database Middleware}
Multi-active database is a middleware that connects the enterprise application with the underlying database. The underlying database could be either a traditional relational database or a NoSQL database. All data definition and data manipulation operations for the database are recorded in an operation log that is maintained using the blockchain technology and is immutable, i.e. it cannot be modified or deleted. The operation log can be used to regenerate the database and therefore can be used for database audit. 
A user application calls the \textsc{ChainSQL} APIs to obtain the transaction data suitable for the blockchain network and sends the transaction data to a network node. The node has to authenticate and validate the newly arrived data before it sends the data to other nodes in the network. The network nodes achieve consensus on the data in the form of transactions that are grouped together as blocks. If consensus is achieved, every node in the network has exactly the same data stored as a set of blocks in order. The nodes configured with the database send the data to the database to synchronise the database operations. 
In case of a node failure, the user can switch to any other node on the network seamlessly. This ensures zero-recovery-time and a multi-active database deployment in real-time. The fault node is restored from the most recent checkpoint during the recovery process. One of the concerns in a multi-active database is the security of data when it is being transmitted across the network. The middleware provides both symmetric and asymmetric encryption schemes from which the user can choose the appropriate security mechanism for the data. Another important aspect of the multi-active middleware is the expandability of the blockchain network nodes. A new node can automatically get the log from an existing node in the network and can replay that log to generate its own version of the database which is the same as other network nodes. Once it is established that the new node has the same data as the other network nodes, it can also participate in consensus build-up and synchronous data writing. 

\subsection{Multi-disaster Recovery Middleware}
Multi-disaster Recovery is also a middleware that connects the database production nodes of the enterprise application with the disaster recovery nodes. 
As mentioned earlier, the user operations are recorded as log files, e.g. Binlog, Redo Log and so forth, in the production centre and are analysed prior to a blockchain transaction generation. During disaster recovery, the first step is to achieve the consensus for the blockchain network that must include the backup node such that the backup node has exactly the same data as every node on the blockchain network. When a new block is generated, the backup node reads the block and sends it to the disaster recovery centre. The recovery centre performs the database backup using the transaction data. Thus, if a node fails at the production centre, the users switch to the recovery centre to complete the task. This is achieved by elevating a backup node to the status of a production node. 
The data of the production centre is transmitted to the disaster recovery centre within ten (10) seconds of a node failure and the log is immediately re-executed to achieve the recovery point objective within a user-specified time.  

\textit{Application.} An important application scenario where \textsc{ChainSQL} has been used is in a banking environment. The business requires the core business data to be protected and the most recent data to be available across the whole business process. The encryption based tamper-resistance feature of the \textsc{ChainSQL} along with the multi-active database ensures security of the customer data and continuous business operations. In case of a node failure, the data-level disaster recovery backup system activates seamlessly thus ensuring the continuity of the business operations. 


\section{Related Work}
%
%
Many blockchain systems have been proposed in literature of which Bitcoin,  Ethereum and Ripple are some of the notable blockchain systems. The study~\cite{DBLP:journals/corr/abs-1708-05665} is a detailed note on data processing for blockchain systems. A number of initiatives have been taken from the data storage and retrieval perspective.  BigChainDB~\cite{mcconaghy2016bigchaindb}, for instance,  combines NoSQL document-based database capabilities for fast queries and reliability of a blockchain. The tamper-resistance is achieved via shared replication, reversion of disallowed updates or deletes, and cryptographic signing of all transactions. However, BigChainDB only supports MongoDB and lacks support for SQL databases. Another notable blockchain system, Ustore~\cite{dinh2017ustore}, uses locality-aware partitioning and remote direct memory access to achieve fast retrievability but in-memory data storage in Ustore compromises the disaster-recovery capability of the system.
%

\section{Conclusion}
In this proposal, we demonstrate the key features of \textsc{ChainSQL} and its novel applications through two usecases that are implemented as a middleware between the user application and the database. The first usecase is a tamper-resistant multi-active database and the second is a data-level disaster recovery backup. \textsc{ChainSQL} is the first system of its kind that features the tamper-resistance of the blockchain and the fast query processing of the distributed databases. The utility of the \textsc{ChainSQL} is evident from its business usecases in domains including finance and supplychain and therefore, it offers promising application scenarios for future. Future research on spatial data, dynamics, data analytics, sharding, and verification may be conducted based on the system~\cite{DBLP:conf/er/CaoCCJQSWY12,DBLP:conf/jcai/WangLQZZ09,DBLP:conf/dasfaa/QuZYHYL11,DBLP:journals/jetai/ZhouQT17,DBLP:conf/edbt/NobariQJ17,DBLP:conf/fit/Muzammal16,DBLP:journals/kais/MuzammalR15,DBLP:conf/pakdd/MuzammalR11,DBLP:conf/pkdd/QuLJZF14,DBLP:conf/waim/TanZQL14,DBLP:conf/wise/NurgalievMQ18}. 
%
%

\section{Acknowledgements}
The work was partially supported by the CAS Pioneer Hundred Talents Program, China [grant number Y84402, 2017], SIAT-Peersafe IoT Security Lab supported by PeerSafe and PeerCome in Beijing and Shenzhen, China, respectively [grant number Y7Z0181001, 2017], and CAS President’s International Fellowship Initiative , China [grant number 2018VTB0005, 2018]. The authors would also like to acknowledge the application development contributions made by Xiaoming Lu and other developers at Peersafe and SIAT.

\bibliographystyle{unsrt}
\bibliography{sigproc}  
%
%
\end{document}